\documentclass{XrU2005}
\usepackage{epsfig}

\title{First Results from the Swift UVOT}
\author{K. O. Mason}
\affil{Particle Physics and Astronomy Research Council, Polaris House, North Star Ave, Swindon, Wilts SN2 1SZ, UK}
\author{A. J. Blustin}
\affil{Mullard Space Science Laboratory, University College London, Holmbury St. Mary, Dorking, Surrey, RH5 6NT UK}
\author{P. W. A. Roming}
\affil{Department of Astronomy and Astrophysics, Pennsylvania State University, 525 Davey Laboratory, University Park, PA 16802, USA}

\begin{document}


\maketitle

\begin{abstract}
We review some of the initial data from the UVOT telescope on the Swift observatory. Statistics
based on about six months of data suggest a dark burst fraction of about 50\% when combining
both UVOT and ground-based observations. There is evidence that some bursts have a large
gamma-ray efficiency, which may be due to strong magnetic fields in their ejecta. The bright
GRB050525A shows behaviour broadly consistent with expectations from the simple fireball
model for bursts, including evidence for a reverse shock component in the UVOT data, and
an achromatic break in decay slope indicative of a jet break. Other bursts observed with
Swift have a shallow decay initially which is difficult to reconcile with the simple model.
Replenishment of the forward shock energy by continued ejection of material from the
central engine, or initial injection of material with a range of velocities, offers a potential
explanation. In the case of the XRF050406 an initially rising optical afterglow flux
followed by a shallow decay may be due to observation of a structured jet from a significant
off-axis angle.
\end{abstract}

\section{Introduction}

%
The Swift  observatory, launched in November 2004, is breaking new ground in the 
study of Gamma-ray Bursts (GRB). It is able to rapidly locate new bursts in 
its 1.4 sterradian field-of-view Burst Alert Telescope (BAT; Barthelmy et al. 2005) and slew to 
bring its narrow field X-ray Telescope (XRT; Burrows et al. 2005) and UV/Optical Telescope (UVOT; Roming et al. 2005a)
to bear on that location in the sky within about 1 minute. At the same time, information on burst location and
properties is immediately communicated to the ground where it is disseminated to observers world-wide via the
Gamma-ray bursts Coordinates Network (GCN).

In this paper we review some early GRB results from Swift, highlighting the contribution made with data from the UVOT.

\section{Detection Statistics}

It is well known that not all GRB have detectable optical afterglows. However, the proportion of these
so-called 'dark bursts' is debated, with estimates based on Beppo-Sax data suggesting that they comprise
about 50\% of the total burst population (e.g. De Pasquale et al. 2003) while HETE-II data suggest that 
less than 10\% of bursts are optically dark (Lamb et al 2004). Possible explanations for dark bursts include
a high redshift (Bromm \& Loeb 2002; Fruchter 1999), absorption in a dense circumburst medium (Lazzati et al. 2002), 
intrinsic faintness (De Pasquale et al. 2003; Roming et al. 2005b) or a rapidly declining afterglow (Groot et al. 1998).
\begin{figure}
\epsfig{file=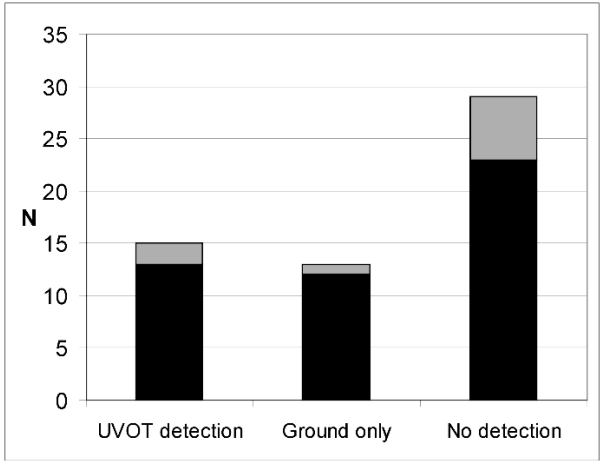,width=1\linewidth}

\caption{Detection statistics for Swift bursts between 24 Jan 2005 and 22 Sep 2005.
Shown are the number of bursts detected with UVOT (column 1), detected with ground-based
telescopes but not UVOT (column 2) and not detected in the optical/IR by any means (column 3).
For each category, we distinguish the number of bursts that were observed with Swift
within 1 hour of the BAT trigger (black area) from those that were not observed until
greater than 1 hour after the trigger (grey area).}
\end{figure}

\begin{figure}
\epsfig{file=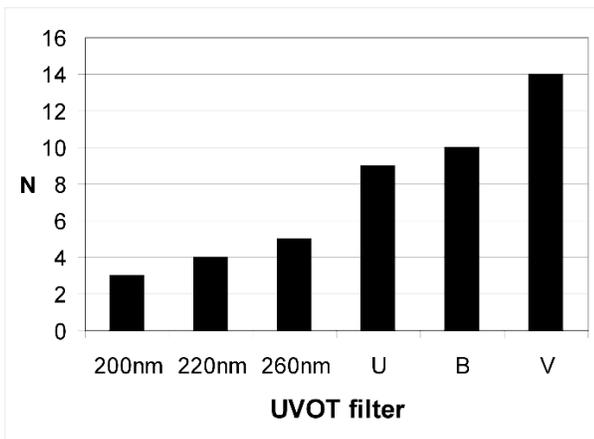,width=1\linewidth}

\caption{Detection statistics for UVOT-detected bursts between 24 Jan 2005 and 22 Sep 2005
as a function of UVOT filter.}
\end{figure}

The Swift data offers the advantage of a sample of bursts that has been uniformly observed very soon after the initial trigger. 
Fig.~1 shows statistics on bursts observed by Swift using UVOT between 2005 Jan 24 (when UVOT was commissioned) and 2005 Sep 22. 
We distinguish between the (majority of) bursts that were observed within an hour of the trigger (and usually 
within a few minutes) and those that were not observed until more than 1 hour after the trigger. A delay in
slewing to a new burst can occur, for example, because the burst occured in a region of the sky where pointing of
the spacecraft is constrained. We divide the bursts into those that were detected with UVOT, those that were detected
using ground-based telescopes only (which usually means that the burst was too red to be detected with UVOT, which has
a long wavelength cut-off of about 650nm), and those that were not detected in the optical/IR by any means.  
These simple statistics suggest that the dark burst fraction among the Swift sample is 48\% for the sample of
48 bursts that were observed using UVOT within one hour. 

\begin{figure}
\epsfig{file=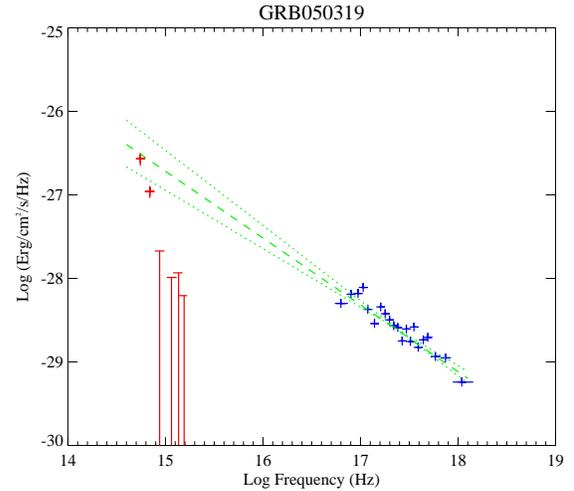,width=1.0\linewidth}

\caption{The multiwavelength spectrum of GRB050319 plotted as F$\nu$ vs log frequency in the observer frame, 
and averaged over the interval between 240s and 930s after the burst trigger. Measurements with the UVOT
taken through its six broad-band filters, centered on 200nm, 220nm, 260nm, U, B abd V,  are shown together with the 
spectral distribution inferred by fitting the count-rate as a function of energy recorded with the XRT.
The best power law fit to the XRT data is illustrated as the dashed line. The 1$\sigma$ bounds on the slope are
indicated by the dotted lines.
The XRT flux points have been corrected for
Galactic absorption equivalent to N$_{\rm H} = 1.17 \times 10^{20}$ cm$^{-2}$.
}
\end{figure}

Fig.~2 shows the detection statistics for bursts as a function of UVOT filter. Unsurprisingly, the highest rate of
detection occurs in the reddest filter and the detection frequency declines monotonically towards the blue. This
presumably reflects the distribution of dust reddening and/or redshift amongst the bursts. Reddening will occur due to dust,
either in the rest frame of the burst or in our Galaxy, while the effects of rest-frame dust reddening and absorption
due to the Lyman edge, or the Lyman forest, will be seen at progressively 
longer wavelengths as the redshift increases. An example is shown in Fig.~3, which shows the combined X-ray and UV/Optical
spectrum of GRB050319 measured using Swift during a specific time interval (Mason et al. 2005). 
The V-band measurement taken with UVOT lies on an
extension of the power-law that models the X-ray spectrum. However, the B-band detection,
and the upper limits to the flux measured in the UVOT filters blueward of B, all lie substantially below the extrapolated
power law. Interpreting this deficit as being due to the Lyman absorption edge redshifted into the UVOT band
suggests a redshift for the burst of about 3.8. In fact, Fynbo et al. (2005) report an absorption line system in the
source at a redshift of 3.24 which is probably the host galaxy. This is consistent with the Swift broad-band spectrum
if there is also significant line opacity in the spectrum, due for example to the Lyman forest.

Roming et al. (2005b) have discussed the detectability of GRB in more detail. They consider the X-ray and optical
flux of GRB afterglows at a set time, 1 hour, after the burst, and compare these with the Gamma-ray fluence. 
The Gamma-ray fluence is a natural measure of the radiated energy of the GRB, while the X-ray flux is a proxy for
the kinetic energy of the fireball's blastwave. They find that there is a large spread in the ratio of Gamma-ray fluence
to X-ray flux one hour after the burst. They highlight three bursts in particular, GRB050223, GRB050421, and GRB050422, which
have a high Gamma-ray to X-ray ratio, none of which are detected in the optical band. This implies Gamma-ray 
efficiencies as high as 90\%, which is difficult to account for in the standard fireball model. They suggest that
the flows in these bursts may be highly magnetised, and that a large fraction of the energy in the ejecta is locked up 
in the magnetic field, at least in the early phases of the expansion.

\section{The bright burst GRB050525A}

GRB050525A was a relatively bright burst at a spectroscopic redshift of 0.61 that was followed by both the
XRT and UVOT instruments from soon after the BAT trigger (Blustin et al. 2005). The data on the X-ray and optical
afterglow decay are shown in
Fig.~4. The data taken through the various UVOT filters are normalised together in this plot (there is no
evidence of a colour dependent decay), while the relative normalisation of the X-ray data is arbitrary
and chosen for display clarity.
 
\begin{figure*}
\epsfig{file=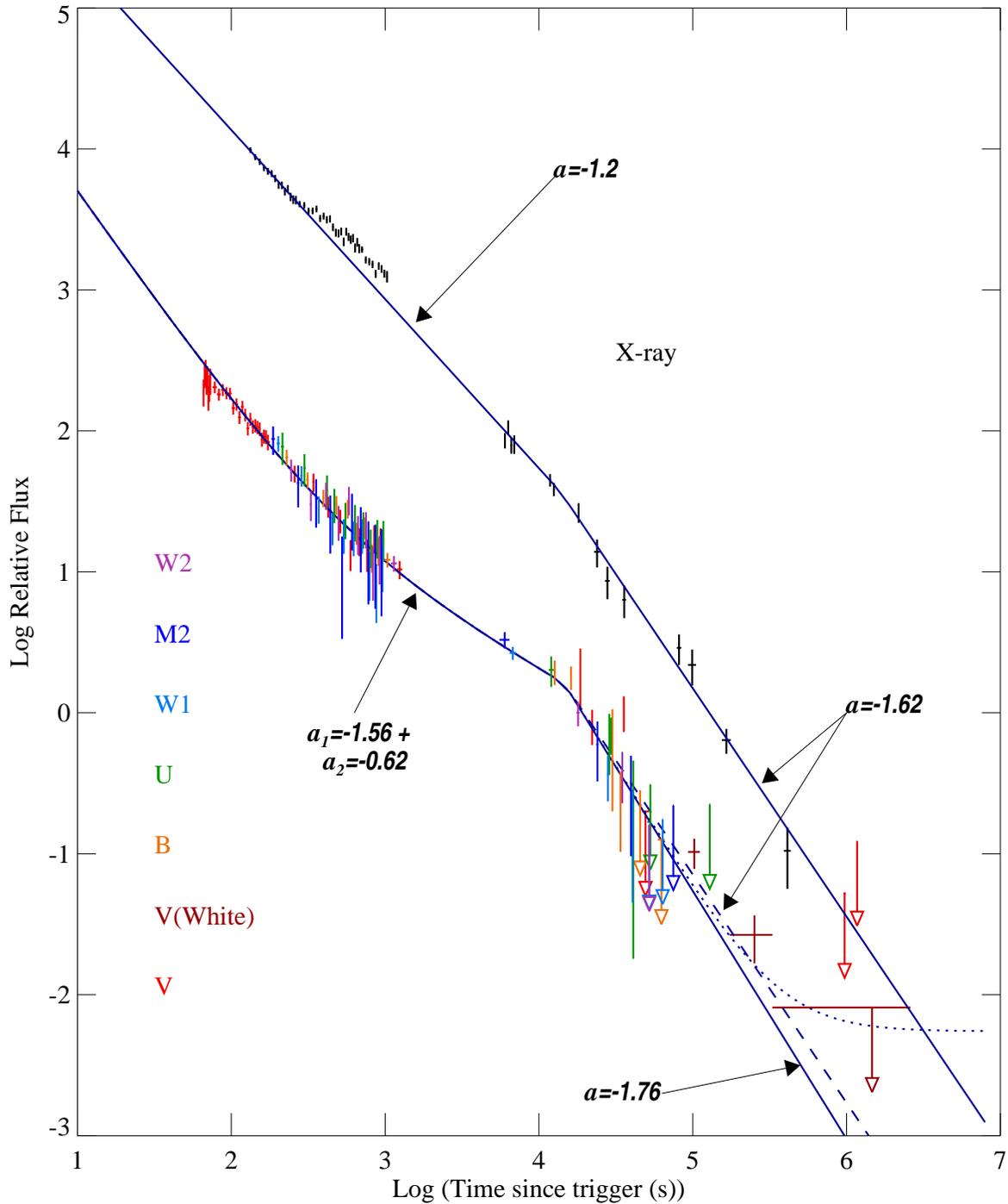,width=1.0\linewidth}

\caption{Comparison of flux decay of GRB050525A in the X-ray and UVOT bands. The UVOT data through different filters have been normalised 
in the interval up to T+1000s, and the data taken through the different filters are plotted together. 
The relative normalisation of the X-ray and optical/UV data is arbitrary. The best fit
broken power law model is plotted through the X-ray data. The best fit double power law with break 
is plotted through the UVOT data (see text). The dashed line has the same post break slope as the X-ray data. The dotted line
is the best fit model with a constant flux added corresponding to the value measured by 
Soderberg et al. (2005) using HST/ACS.}
\end{figure*}

The X-ray afterglow of GRB050525A decays initially with a power-law slope ($\alpha$) of $-1.2$. After about 300s, the 
X-ray flux exhibits an excess with respect to this power-law, which persists until there is a gap in coverage
due to Earth-occultation of the source. After the occultation gap the X-ray data again lie on the original power-law,
suggesting that the excess flux was part of a relatively short-lived flare. There is a break in the power-law after
about $10^4$s to a new, steeper slope of $-1.62$.

The optical decay has a distinctly different form. It is initially steeper than the X-ray curve, before flattening
to a shallower slope. Once again, after about $10^4$s the slope steepens to a value that is consistent with
the X-ray data in the same time interval. The optical decay before the $10^4$s break can be represented by a 
combination of two power-laws. This fits naturally with the idea that there is an initial steep drop due to the fading of
a reverse shock component, which is only seen in the optical band, combined with a flatter decay component from the forward
shock. The behaviour of the source is clarified when one looks at the behaviour of the multiwavelength spectrum
with time. This is shown in Fig.~5, which shows the X-ray and optical/UV spectrum of GRB050525A at three epochs
in the decay, 250s, 800s and 25000s after the BAT trigger. A single power-law spectrum is consistent with both the
X-ray and optical/UV data at 25000s after the burst, but not at the earlier epochs, where the optical/UV flux is
suppressed relative to an extrapolation of the X-ray spectrum. The `recovery' of the optical/UV emission relative to the
X-ray flux in both the spectral and time domain suggests that we might be seeing the migration of the synchrotron 
cooling frequency through the optical/UV band. This migration is somewhat faster the prediction $\nu_c\propto t^{0.5}$ of the
simple fireball model (e.g. Zhang \& M\'esz\'aros 2004), possibly due again to the liberation of 
energy locked up in magnetic fields. The sense of the spectral evolution favours expansion into a constant density
(interstellar) medium rather than the $1/r^2$ density dependance of a stellar wind.

The best-fit slope of the pre-break forward shock component is somewhat shallower than the $\alpha=-0.9$ expected from
a simple fireball model. However the fitted value is sensitive to the exact form of the 'reverse shock' component.
We also note that Klotz et al (2005) suggested that the optical decay suffered a `re-brightening' episode about 2000s
after the burst,
during the gap in UVOT coverage. If we include such a re-brightening in our model fits, the overall forward shock
decay slope steepens to about $\alpha=-1$, though the quality of the fit is marginally worse. The fit parameters 
for the smooth (`best fit') and re-brightening (`step fit') models, together with the 
predictions of the fireball model are summarised in Table~1. The fitted slope of the
reverse shock component is also sensitive to whether we include a re-brightening, and ranges between $\alpha=-1.5$ without
re-brightening to $\alpha=-2.1$ with.

\begin{figure}
\epsfig{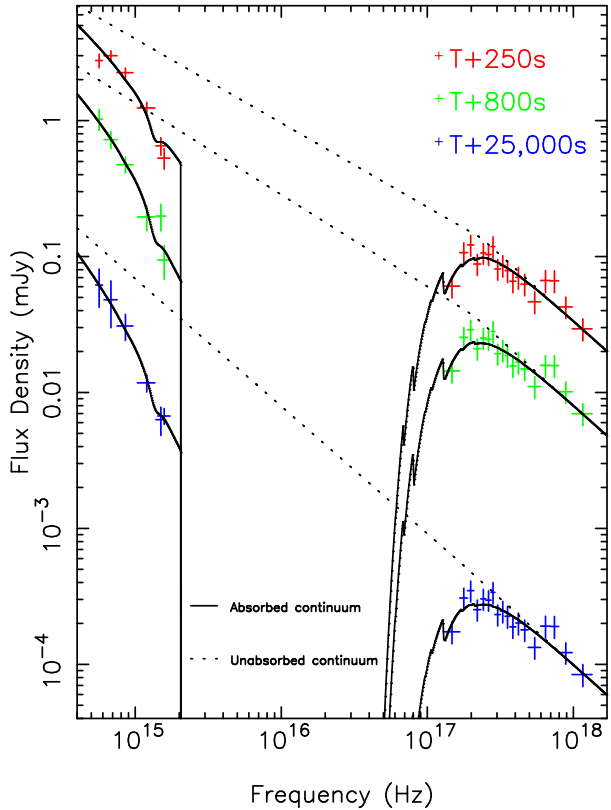}

\caption{    UVOT and XRT data on GRB050525A interpolated to the epochs T+250s, T+800s and
T+25,000s, together with best spectral fit models (solid lines). The dotted lines
represent the intrinsic continuum of the source, before extinction and absorption 
from gas and dust in both the Milky Way and the host galaxy.
}
\end{figure}

The steepening of both the X-ray and optical/UV decay slope after about $10^4$s can be interpreted as a jet break. This
is supported by the fact that the break is achromatic, i.e. the break occurs at the same time in both the X-ray and 
optical/UV range, and
the post-break slope is consistent in the two bands. The best fit to a single broken power-law model yields a
break time of about 14000s. However the post-break slope ($\alpha=-1.6$) is shallower than the $\alpha=-2.2$ predicted
by simple fireball models. This could be due to the details of how the jet evolves, or the break might actually be
more gradual than the simple model would suggest. In this case a slope of $\alpha=-2.2$ is reached at a later time. 
Such a model
could be consistent with the data (but is not required) and  
yields an effective break time at
about 50,000s-60,000s. A break at 14000s, combined with the measured isotropic-equivalent energy emitted 
in the burst, suggests
a jet opening angle of about $3.2^\circ$, assuming a uniform jet. The opening angle increases to about $5^\circ$ if we 
adopt the later
time implied by a gradual break. 

\begin{table}
  \begin{center}
    \caption{Comparison of model fits to the GRB050525A data with expectations from the fireball model.}\vspace{1em}
    \renewcommand{\arraystretch}{1.2}
    \begin{tabular}[h]{lrcc}
      \hline
      &  Fireball  &  Best fit &  Step fit\\
      &   model$^1$ &  &  \\
      \hline
      X-ray decay           & -1.15 &\multispan2{ -1.2}  \\
      Optical decay$^2$         & -0.9 & -0.62 & -1.04 \\
      X-ray spectral slope  & -1.1 &\multispan2{ -0.97}  \\
      Optical spectral slope& -0.6 &\multispan2{ -0.60} \\
      \hline \\
      \end{tabular}
    \label{tab:table}
  \end{center}
$^1$ For p=2.2, ISM slow cooling model\\
$^2$ Forward shock
\end{table}

In all, the properties of GRB050525A show good agreement with expectations based on the standard fireball model
(Zhang \& M\'esz\'aros 2004). There is evidence for a reverse shock component in the optical/UV decay curve, and
for migration of the synchrotron cooling frequency through the optical/UV band. There is also evidence for a light
curve `jet' break, which is expected when the fireball Lorentz factor decreases to the point where the beaming angle
of the emitted radiation exceeds the collimation angle of the jet (Rhoads 1999; Sari, Piran \& Halpern 1999). 

\section{GRB Diversity}

GRB050525A may not be a typical Swift burst in behaving broadly in line with the standard fireball model.
In the case of GRB050319, the decay of both the optical and X-ray flux is much shallower than predicted, with
a slope of about $\alpha=-0.5$ (Mason et al. 2005; Cusumano et al. 2005). This behaviour persists until at 
least $3\times10^4$s after the burst, before the X-ray slope, at least, steepens to a value of $\alpha=-1.1$
(Cusumano et al. 2005). This behaviour is not unique. For example De Pasquale et al. (2005a) find a similar
initial slope in the X-ray decay of GRB050401, persisting for a few thousand seconds after the burst, before
steepening  to a value $\alpha=-1.5$. One explanation for this phenomenon is that the central engine
continues to inject energy into the afterglow, at a decreasing rate, for some time after the initial
burst (Zhang \& M\'esz\'aros 2004). The decay rate steepens once the injection has ceased. A similar effect
can be produced by ejecta that has a range of Lorentz factors, with the shock being `refreshed' as it decelerates
by initially slower moving shells that catch up with it (see De Pasquale et al 2005a, and Stanek et al. 2001, 
Bjornsson et al. 2002 in the context of GRB010222). Even more extreme behaviour is seen in GRB050712 
(De Pasquale et al. 2005b) where the optical flux is flat, or even rises, during the first few hundred seconds
following the burst. This is a case where the XRT data shows continued flaring during the same interval.

Another interesting case is the Swift data on the X-ray flash XRF050406 (Schady et al. 2005), which represent the earliest
observations yet made of the optical emission of an X-ray flash, starting 88s after the BAT trigger. The
optical emission is faint, but consistent with a rising flux in the first 200s of the afterglow, decaying
thereafter with a shallow slope $\alpha\sim -0.7$. Schady et al. argue that both the soft X-ray spectrum of the
initial burst emission, and the initially rising flux and shallow decay are consistent with observation of
a structured jet viewed slightly off axis. In this case the Lorentz-beamed emission of the main jet core is not
within the line of sight when the burst first goes off, and we see only fainter and softer emission from the outer
portions of the jet. As the jet core decelerates, the beaming angle widens and we seen enhanced emission along our
line of sight.

\section{Conclusions}

The Swift observatory is gathering unique data on the prompt and early afterglow emission of GRB. Data from the UVOT
are providing an unprecedented glimpse of the early optical afterglow emission, which can be combined with the
X-ray data taken simultaneously with XRT to study the behaviour of the afterglow across a range of frequencies
and constrain physical models.
Multi-filter data from the UVOT provides a valuable indicator of redshift, to supplement ground-based spectroscopy. 
The Swift data have already revealed considerable diversity in behaviour among GRB, and we look forward to
building up larger samples as the mission progresses, with which to investigate the full range of GRB phenomenology.

%
%
%
%


\section*{Acknowledgments}

We are greatful to the entire Swift team for their excellent and dedicated work in 
operating this challenging mission.

\end{document}